\documentclass[10pt,twocolumn,pre,aps,superscriptaddres,floatfix]{revtex4-1}

\usepackage{amssymb}
\usepackage{amsmath}
\usepackage{bm}

\usepackage[linkcolor = blue, citecolor = blue, urlcolor = blue, colorlinks = true]{hyperref}

\usepackage{graphicx}
\usepackage{wrapfig}
\usepackage[dvipsnames]{xcolor}

\usepackage[utf8]{inputenc}
\usepackage{lipsum}

\begin{document}

\title{Hydrodynamic Lubrication in Colloidal Gels}

\author{K.W. Torre}
 \email{k.w.torre@uu.nl}
\affiliation{
 Institute for Theoretical Physics, Center for Extreme Matter and Emergent Phenomena, Utrecht University, Princetonplein 5, 3584 CC Utrecht, The Netherlands 
}
\author{J. de Graaf}
\affiliation{
 Institute for Theoretical Physics, Center for Extreme Matter and Emergent Phenomena, Utrecht University, Princetonplein 5, 3584 CC Utrecht, The Netherlands 
}

\date{\today}


\begin{abstract}
Colloidal gels are elasto-plastic materials composed of an out-of-equilibrium, self-assembled network of micron-sized (solid) particles suspended in a fluid. Recent work has shown that far-field hydrodynamic interactions do not change gel structure, only the rate at which the network forms and ages. However, during gel formation, the interplay between short-ranged attractions leading to gelation and equally short-ranged hydrodynamic lubrication interactions remains poorly understood. Here, we therefore study gelation using a range of hydrodynamic descriptions: from single-body (Brownian Dynamics), to pairwise (Rotne-Prager-Yamakawa), to (non-)lubrication-corrected many-body (Stokesian Dynamics). We confirm the current understanding informed by simulations accurate in the far-field. Yet, we find that accounting for lubrication can strongly impact structure at low colloid volume fraction. Counterintuitively, strongly dissipative lubrication interactions also accelerate the aging of a gel, irrespective of colloid volume fraction. Both elements can be explained by lubrication forces facilitating collective dynamics and therefore phase-separation. Our findings indicate that despite the computational cost, lubricated hydrodynamic modeling with many-body far-field interactions is needed to accurately capture the evolution of the gel structure.
\end{abstract}

\maketitle


\section{\label{sec:intro}Introduction}

Colloidal particles dispersed in a fluid medium can form a gel, when there are short-ranged attractions between the colloids with an interaction strength that significantly exceeds the thermal energy $k_{\mathrm{B}} T$~\cite{lekkerkerker1992poon, poon2002physics, bergenholtz2003gelation, chen2004microscopic}; here, $k_{\mathrm{B}}$ is the Boltzmann constant and $T$ the temperature. These interactions can be induced by,~\textit{e.g.}, the presence of polymers~\cite{asakura1958interaction} (depletion) or van-der-Waals interactions~\cite{Royall2021, hartley1985dispersion} (solvation). The depth and short-ranged nature of the attractive potential well interfere with thermal rearrangement of clustered colloids. This arrests the system’s natural tendency to fully phase separate~\cite{foffi2002evidence, zaccarelli2009colloidal}. Diffusion-based aggregation together with this arrested cluster dynamics leads to the formation of an open, space-spanning network structure~\cite{carpineti1992spinodal, zaccarelli2007colloidal, Royall2021}. The resulting material possesses useful properties,~\textit{e.g.}, it can support the gel’s buoyant weight against gravity for a finite time~\cite{buscall1987consolidation, allain1995aggregation, allain2001systematic, starrs2002collapse, weitz2005gravitational, bartlett2012sudden, zaccarelliHarich2016, padmanabhan2018gravitational}, behaving solid-like during this period, yet flow under moderate applied stresses~\cite{koumakis2015tuning}, behaving liquid-like. This ability to yield easily and reform~\cite{smith2007yielding} has led to the widespread use of particle gels in industrial, medical, and academic settings, for example, care products, printing inks, foodstuffs, crop protection, and pharmaceutical suspension formulations~\cite{larson1999structure, eryt2022, food-soft-materials2005, crop_protection2006}.

Gels are intrinsically out of equilibrium and therefore age after they form~\cite{verhaegh1999transient, cipelletti2000universal, manley2005time}. A gel's structure and its mechanical response are intimately linked~\cite{shah2003microstructure, koumakis2015tuning, tsurusawa2019direct}. This necessitates understanding of formation and aging in order to begin to predict and ultimately control the properties of colloidal gels. Both processes have been suggested to depend (strongly) on the nature of the fluid medium, in which the colloids are suspended~\cite{furukawa2010key, whitmer2011influence, vargaswan2015, royall2015probing, varga2016hydrodynamic, gelhydroJoost2019}. That is, unlike equilibrium systems, for which the particle dynamics are known not to affect the average arrangement of colloids, hydrodynamic interactions (HIs) can affect gel structure by impacting the kinetics of formation. HIs result from momentum conservation in the fluid, which gives rise to long-ranged forces --- decaying with the inverse of the center-to-center separation $1/r$ --- between moving suspended particles~\cite{happel1983mechanics, kim2013microhydrodynamics}. The forces inducing HIs can be externally applied, stem from the interaction potential, or are thermal in nature. The latter two play a prominent role in gel formation (without gravity), thus it is relevant to quantify their effect on structure when accounting for the suspending fluid.

Over the past two decades, various research groups have performed numerical studies of colloidal gels that account for HIs~\cite{tanaka2000simulation, yamamoto2008role, furukawa2010key, whitmer2011influence, vargaswan2015, royall2015probing, varga2016hydrodynamic, zaccarelliHarich2016, padmanabhan2018gravitational, varga2018normal, varga2018modelling, swan-furst2019, gelhydroJoost2019,turetta2022role,joost2023hydrodynamic}. Introducing HIs into particle simulations poses a significant challenge. In the friction-dominated regime appropriate to colloidal hydrodynamics, a Newtonian fluid satisfies the Stokes equations. Exactly solving the full set of Stokes equations, for which the particles act as boundary conditions, is practically infeasible for all but the simplest systems~\cite{happel1983mechanics,kim2013microhydrodynamics}. Additionally, HIs decay as $1/r$, which makes the dynamics in a particle suspension an intrinsically many-body problem~\cite{Brinkman1949ACO, howells1974drag}. A number of approximating techniques have been proposed over the years to make progress~\cite{dunweg2009lattice, dickinson2013structure, bolintineanu2014particle}. These techniques weight the accuracy, by which HIs are approximated, against computational efficiency. Prominent examples of methods that have been used in studying colloidal gels include: Fluid Particle Dynamics (FPD), Lattice-Boltzmann (LB), and Stokesian Dynamics (SD). In view of the approximating nature of gel simulations with HIs, it is not surprising that many different (seemingly conflicting) results on the impact of HIs have been reported.

De Graaf~\textit{et al.}~\cite{gelhydroJoost2019} recently used lattice-Boltzmann (LB) method~\cite{dunweg2009lattice,rohm2012lattice} to account for the many-body, far-field interactions between gelling colloids. These authors, revealed two distinct regimes in their simulations with and without HIs, and a crossover at a colloid volume fraction of $\phi \approx 0.16$. For $\phi < 0.16$, HIs were found to accelerate gelation, while for greater $\phi$, the effect was seen to be reversed. The apparent substantial differences in the gel structure, when comparing gels with and without HIs at equivalent time, particularly at low $\phi$, were found to become negligible, when comparing the systems at equivalent ``structural times''~\cite{gelhydroJoost2019}. The findings of De Graaf~\textit{et al.}~appear to reconcile the contradictory observations reported in previous simulation studies with HIs, requiring only limited reference to the various approximations made by other authors.

However, similar to the vast majority of simulation studies, the authors of Ref.~\cite{gelhydroJoost2019} ignored hydrodynamic lubrication forces (HLFs). These forces arise when colloidal particles are separated by less than $\approx 10\%$ of their particle diameter and are highly dissipative~\cite{jeffrey1984calculation}. Motion parallel to the line connecting the centers of two \textit{spherical} colloids is associated with a hydrodynamic friction that diverges as $1/h$, where $h$ measures the surface-to-surface separation. This divergence results from the force required to squeeze the fluid out of the small gap between the colloids. Such an interaction has the potential to interfere with gelation, as the attractions induced by depletion interactions also typically fall in this separation regime. Analogous divergences exist for the other modes of relative displacement,~\textit{e.g.}, there is a logarithmic divergence for motion orthogonal to this connecting line.

In the context of colloidal gels, the work of Bybee and Higdon~\cite{bybee2009hydrodynamic} is often referenced to justify neglecting HLFs. These authors approximated HIs by considering exclusively HLFs between nearly touching particles. Their study revealed that the percolation line and microstructure of the gel did not exhibit noticeable changes, when compared to the results obtained using non-lubricated Brownian Dynamics (BD). Thus, these authors argued that near-field HIs contribute only minimally on the time scale of gelation. However, the recent work by Townsend~\textit{et al.}~\cite{townsend2018anomalous} on HIs in colloidal suspensions, demonstrated that neglecting far-field HIs can lead to inaccuracies, even when colloidal particles are in close proximity to each other.

An additional argument used to neglect HLFs is that in many experimental systems it is not clear that the fluid is Newtonian. That is, macromolecules (polymers) may have been added to induce gelation \textit{via} depletion, which could affect the medium's hydrodynamic response.  Moreover, even very well synthesized colloidal spheres are not perfectly spherical or hard~\cite{royall2013search}. This has led to debate on the relevance of contact interactions~\cite{immink2019using,wang2019surface,nguyen2020computer,van2022emergence}. Both of these effects limit the applicability of analytic lubrication expressions, which are derived for perfect spheres in a Newtonian solvent. Nonetheless, it is important to fundamentally understand what effect HLFs can have on gelation, gels structure, and aging, before discounting these in favor of other mechanisms. This leads us to revisit colloidal gelation in this work.

It should be noted that LB --- and many other approximations for HIs --- possesses a near-field increase of the hydrodynamic friction between approaching particles~\cite{gelhydroJoost2019}. In the case of embedded particles that couple to the lattice fluid on a sub-grid scale, the increase is non-divergent and a mix between far-field flows and compressibility / approximation-level artifacts. For Ladd-type boundaries, there are also issues, but these can be explicitly lubrication corrected~\cite{nguyen2002lubrication},~\textit{i.e.}, it is subtracted and replaced by accurate analytical lubrication approximations. However, this it is not clear how to effectively do this when studying the dynamics of suspended sub-grid particles experiencing thermal fluctuations. Lubrication corrected Ladd-type particles typically require too many lattice sites to effectively study gelation, as the number of particles is limited by the computational requirements. A different approach is thus needed.

Here, we thus consider the effect of truncating the HIs at four distinct levels, which allows us to chart the specific effect of each approximation. The four levels of approximation are: (i)~Free-draining spheres (FD) --- effectively captured by BD simulations --- will serve as our reference point. This model considers forces between the fluid and colloids only at the one-body level,~\textit{i.e.}, the colloids experience Stokes drag but have no long-ranged or short-ranged HIs. (ii)~Rotne-Prager-Yamakawa (RPY) hydrodynamics~\cite{rotne1969variational, swan2016rapid, fiore2017rapid, Fiore2018, pelaez2022complex}, which approximate HIs using far-field Green's functions at a pair level. That is, RPY HIs ignore the intrinsic many-body effects present between suspended spheres, which is typically considered reasonable for low $\phi$. Here, we do not make use of lubrication corrections. (iii)~Stokesian Dynamics (SD)~\cite{brady1988stokesian, phung1996stokesian, sierou2001accelerated, banchio2003accelerated, wang2016spectral, fiore2019fast} without lubrication corrections (SD$^{\textit{ff}}$). This approach accounts for the many-body effects, but has limited accuracy for spheres that approach each other closely, as is the case in colloidal gels. Of the approaches, it is most closely comparable to the work by De Graaf~\textit{et al.}~\cite{gelhydroJoost2019} --- the LB method is many body in nature and we will contrast Ref.~\cite{gelhydroJoost2019}'s results to those obtained here. SD is based on matrix inversion and typically more costly than LB, limiting the number of particles that we studied to about 7,000. (iv)~Lastly, we consider lubrication-corrected SD (simply SD throughout). This is the most accurate of all techniques in describing the interactions between spheres suspended in a Stokes fluid. However, this comes at the price of being the most computationally expensive. Even full SD is approximating in the mid-field,~\textit{i.e.}, when transitioning from lubrication to the far-field expressions.

Using the above four methods, we find that the dynamics of the gel are highly sensitive to the type of hydrodynamic model used, in line with previous studies. However, the (quasi-)steady-state structure remains relatively unaffected between these, provided HLFs are not present. Interestingly, when HLFs are introduced, significant differences are observed, particularly for colloid volume fractions $\phi < 0.138$. In addition to altering the structure of the gel, resulting in smaller voids and clusters at the percolation point, lubrication also accelerates gel aging. This may seem counter intuitive, as lubrication is associated with the strongest hydrodynamic dissipation. The explanation for both these effects is that lubrication interactions facilitate phase separation, by interfering with bonding and prevent the rupture of clusters. Additionally, HLFs suppress non-collective modes in the gel arms, leading to more vigorous dynamics.

The remainder of this paper is organized as follows. In Section~\ref{sec:meth}, we first provide an overview of the numerical methods used in this work. Section~\ref{sec:hydro} contains specific details of the hydrodynamic models considered, while in Section~\ref{sec:character} we present the quantities used to characterize our gel systems. Next, in Section~\ref{sec:results}, we show the results of our simulations and describe the system's dynamics and structure. Here, we also provide the first insights into the effects of HIs and HLFs. In Section~\ref{sec:disc}, we connect our results with the existing literature and provide an explanation for the deviations observed in SD simulations. We also discuss some of the advantages and limitations of the SD method. Finally, Section~\ref{sec:conclusion} concludes with a summary of our findings and an outlook on future directions for particle-based simulations of colloidal gels.

\section{\label{sec:meth}Simulation Setup}

We want to study the influence of hydrodynamic interactions on the structural formation and evolution of a colloidal gel. We do so by performing particle-based simulations of neutrally buoyant spherical particles of diameter $\sigma$, suspended in a fluid with viscosity $\eta$. We simulate only the colloids and account for the presence of the polymers that cause depletion attraction \textit{via} a generalized ``high-exponent'' LJ potential
\begin{align} 
    \label{eqn:LJpotential}
    V_{\mathrm{LJ}}^{\mathrm{he}} &= \epsilon \left [ \left ( \frac{\sigma}{r} \right )^{96} -~2 \left ( \frac{\sigma}{r} \right )^{48} \right ],
\end{align}
where $\epsilon$ the interaction strength is set to $10 k_{\mathrm{B}} T$. This is a smooth approximation to the well-known Asakura-Oosawa-Vrij interaction~\cite{miyazaki2022asakura} when combined with steric repulsion that prevents overlap.

We can write the equations of motion for the entire system of $N$ colloids using three $6N$-dimensional generalised force vectors
\begin{align}
    \label{eqn:eom}
    0 &= \mathcal{F^P} + \mathcal{F^H} + \mathcal{F^B} ,
\end{align}
which group forces and torques together. The first term in the right-hand side of Eq.~\eqref{eqn:eom}, $\mathcal{F^P}$, is a $6N$-dimensional vector with the first $3N$ components containing the inter-particle forces and the last $3N$ entries are zero (our interaction potentials are torque free). The second term, $\mathcal{F^H}$ contains hydrodynamic drag forces. In a steady Stokes' flow, this term can be written as the product of a hydrodynamic resistance matrix, which depends only on the relative distances between particles, and a $6N$-dimensional vector $\mathcal{U}$ containing translational- and rotational-velocities:
\begin{align}
        \label{resmatrix}
        \mathcal{F^H} &= - \mathbf{R}_{\text{FU}} \cdot \mathcal{U}.  
\end{align}
The last term in Eq.~\eqref{eqn:eom}, $\mathcal{F^B}$, contains forces and torques that account for Brownian motion. These are instantaneously correlated through the HIs and their variance is given by the fluctuation–dissipation theorem~\cite{kubo1966, Deutch1971fluct-diss}. In a simulation with time-step size $\Delta t$, we can express this correlation as~\cite{Ermak1978, Bossis1987}:
\begin{align*}
        \langle \mathcal{F^B}(t_i) \mathcal{F^B}(t_j) \rangle - \langle \mathcal{F^B}(t_i)\rangle  \langle \mathcal{F^B}(t_j)\rangle &= \frac{2 k_{\mathrm{B}} T}{\Delta t} \mathbf{R}_{\text{FU}} \delta_{ij},
\end{align*}
where $\delta_{ij}$ represents the Kronecker delta and the angled brackets indicate a time average.

To compute the Brownian contributions, we must specify by which convention we evaluate them~\cite{van1981ito}. For numerical studies, a common choice is the It{\^o} convention. At each (discrete) time $t_i$, we compute $\mathbf{R}_{\text{FU}}$ using the relative particle positions $\mathbf{r}(t_i)$. Following this convention, the Brownian forces acquire a non-zero average
\begin{align*}
        \langle \mathcal{F^B}(t_i) \rangle &= k_{\mathrm{B}} T~\mathbf{R}_{\text{FU}} \cdot \nabla \cdot \mathbf{R}^{-1}_{\text{FU}} ,
\end{align*}
which is usually called ``Brownian drift''. Physically, its presence is required to ensure stationarity under the Gibbs–Boltzmann distribution and generate particle configurations with the correct statistics at equilibrium~\cite{Fiore2018}.

Combining the above, the equation for particle displacement reads
\begin{align*}
    \frac{\Delta \mathbf{r}}{\Delta t} =  \mathbf{R}_{\text{FU}}^{-1} \cdot \mathcal{F^P} + \sqrt{\frac{2k_{\mathrm{B}} T}{\delta t}} \mathbf{R}_{\text{FU}}^{-1/2} \cdot \mathcal{\psi} + k_{\mathrm{B}} T~\nabla \cdot \mathbf{R}_{\text{FU}}^{-1},
\end{align*}
with $\mathcal{\psi}$ a $6N$-dimensional vector containing uniformly distributed random variable with zero mean and unit variance. We must evaluate $\mathbf{R}_{\text{FU}}$ to integrate the colloid trajectories in time. In general, computing an exact hydrodynamic resistance or mobility matrix is a task with a high computational expense, involving the approximate solution of the Stokes equations in the fluid phase surrounding the particles. Section~\ref{sec:hydro} provides details on the four approximations to $\mathbf{R}_{\text{FU}}$, which we used to make it feasible to perform dynamical simulations.

We used periodic, cubic boxes with an edge length $L$ such that the volume fraction of the systems were $\phi \in [0.075, 0.138, 0.225]$. The number of colloids in the boxes was chosen to be $N=$1,000 and $\approx$~7,000, with $L$ suitably modified to achieve a given $\phi$. Unless otherwise explicitly indicated, all simulation results presented use the larger particle number. Following~\cite{gelhydroJoost2019}, we first equilibrated each configuration for $50 \tau_{\mathrm{B}}$ using a purely repulsive inter-particle potential ($r_{\mathrm{cut}} = \sigma$), where $\tau_{\mathrm{B}} = \sigma^{2}/(4D)$ is the Brownian time of the colloids with single-particle translational diffusion coefficient $D$. Subsequently, the gels were formed \textit{via} an instantaneous deep quench from a purely repulsive potential to one with the aforementioned $10 k_{\mathrm{B}}T$ attraction strength, and left to form and age for $[20, 50, 80, 1000]\tau_{\mathrm{B}}$.

The FD, SD$^{\textit{ff}}$, and SD simulations were performed using HOOMD-blue, a GPU-compatible Python package developed in the Glotzer Lab~\cite{ANDERSON2020109363}. For the SD$^{\textit{ff}}$ and SD simulations we used Fiore's external plugin ``Positively-split Ewald'' (PSEv3) to implement the Stokesian Dynamics algorithm~\cite{fiore2019fast}. The RPY simulations were performed using UAMMD, a GPU-accelerated software infrastructure for complex fluids simulations~\cite{pelaez2022complex}. For each data point we ran 3 independent simulations (10 for the smaller system sizes). Each typically took several hours to several days to run on a desktop (i9-10900) with modern GPU (NVidia RTX 2060 Super).

\section{\label{sec:hydro}Hydrodynamic Models}

\begin{figure}[!htb]
\centering
\includegraphics[width=85mm]{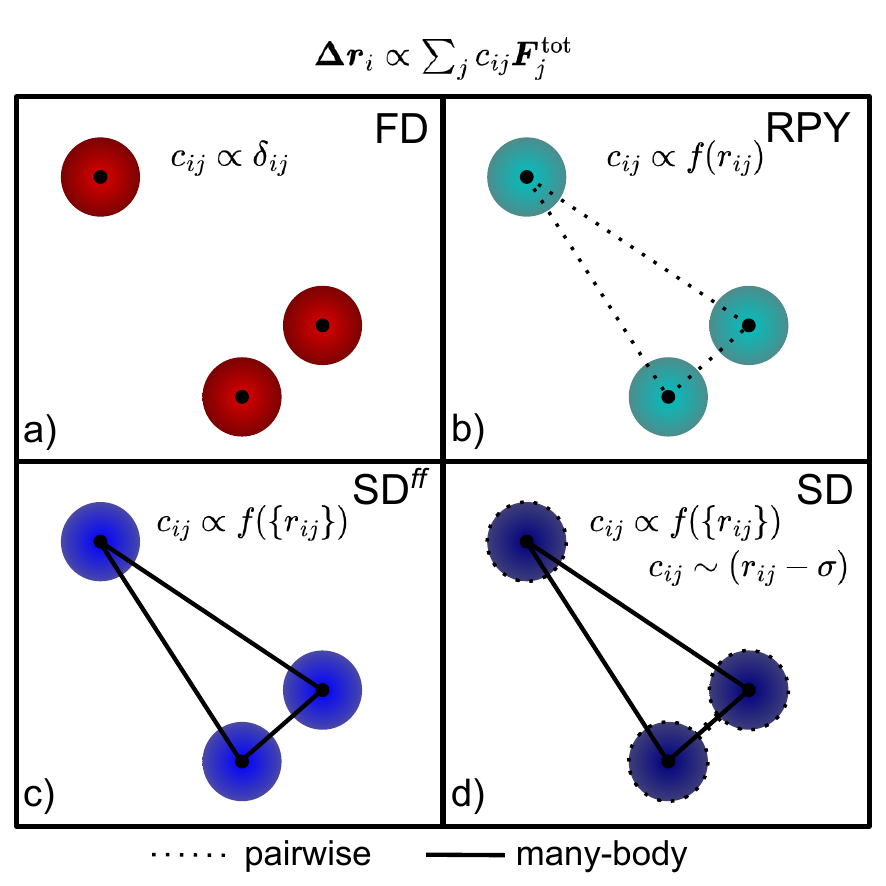}
\caption{\label{fig:panel}Schematic representation of the four hydrodynamic models used. The relation between the displacement of the $i$th particle $\boldsymbol{\Delta r}_{i}$ and the total force applied on it $\boldsymbol{F}^{\mathrm{tot}}_{i}$ and its surrounding particles $\boldsymbol{F}^{\mathrm{tot}}_{j \ne i}$ is indicated at the top; the $c_{ij}$ represent coupling between the particles. (a) In FD simulations, $\boldsymbol{\Delta r}_i$ depends only on the force on the particle itself, $c_{ij} = \delta_{ij}$. (b) In RPY, particles couple pairwise,~\textit{i.e.}, each $c_{ij}$ is a function of the relative distance $r_{ij}$ between the $i$th and $j$th particle, respectively. (c) In lubrication-free SD, SD$^{\textit{ff}}$, there is coupling between all particles, so that each $c_{ij}$ has as argument the set of all distance vectors $\{ r_{ij} \}$. (d) In full SD, short-range lubrication interactions are incorporated, thus the coefficient $c_{ij}$ is modified by an additional pair-wise interaction that causes it to vanish, when the surfaces of particles $i$ and $j$ approach each other (spheres with diameter $\sigma$).}
\end{figure}

In this section, we provide some details on the hydrodynamic approximations used in our work, aimed at contextualizing our results. For each approximation, we provide references to help the interested reader to learn more about that approach. Figure~\ref{fig:panel} gives a visual summary of these approaches and introduces a color coding for the methods that we use throughout.

FD spheres are arguably the simplest model for HIs. Here, HIs are accounted for only at a one-body level. This means that each particle experiences the same drag force (given a velocity), irrespective of the presence of other particles in the simulation volume. In this case, the resistance matrix takes the simple diagonal form $R^{(ij)}_{\text{FU}} = 3 \pi \eta \sigma \delta_{ij}$ and the thermalization is similarly reduced, see Fig.~\ref{fig:panel}a. This makes the FD approximation the fastest method at our disposal; it is fully equivalent to regular BD. The efficiency comes at the cost of neglecting HIs at any relative distance, which means that this approximation produces unrealistic dynamics for suspended colloids. This is not an issue for studying equilibrium systems, though it strongly affects the dynamics of formation~\cite{furukawa2010key, whitmer2011influence, vargaswan2015, royall2015probing, varga2016hydrodynamic, varga2018normal} and collapse~\cite{zaccarelliHarich2016, joost2023hydrodynamic} in colloidal gels.

A more realistic approximation involves a direct construction of the inverse of the resistance matrix, the mobility matrix. This can be done using pairwise Greens' functions for the interactions between spheres, which are called RPY tensors, as depicted in Fig.~\ref{fig:panel}b. At a two-body level, these follow form the combination of Fax{\'e}n's laws for spherical particles and a multi-pole expansion of the solution to Stokes' equation~\cite{rotne1969variational}. For two particles in an unbounded solvent, the tensor takes the form
\begin{align*}
    \mathbf{M}^{(\text{i,j})}_{\text{RPY}} = \frac{1}{3\pi\eta\sigma} 
    \begin{cases} 
        ( \frac{3\sigma}{8r} + \frac{\sigma^3}{16r^3} ) \mathbf{I} + ( \frac{3\sigma}{8r} - \frac{3\sigma^3}{16r^3} ) \mathbf{\hat{r}\hat{r}}, & r  > \sigma, \\ ( 1 - \frac{9r}{16\sigma}) \mathbf{I} + ( \frac{3r}{16\sigma} ) \mathbf{\hat{r}\hat{r}},  & r \le \sigma, \\
        \ \mathbf{I} , & i = j,
    \end{cases}
\end{align*}
where $\mathbf{I}$ is the 3-dimensional identity matrix, $r=|\mathbf{r}_i - \mathbf{r}_j|$, and $\mathbf{\hat{r}}$ is the center-to-center unit vector. It is implied that $\mathbf{\hat{r}\hat{r}}$ represents the Kronecker product of the two vectors and acts as a $3 
\times 3$ matrix. Invoking pair-wise addition, the tensor for a system of $N$ colloids can be written as
\begin{align*}
    \mathbf{M}_{\text{RPY}} = \left( \begin{matrix} 
    \mathbf{M}^{\text{\tiny{(1,1)}}}_{\text{\tiny{RPY}}}&\mathbf{M}^{\text{\tiny{(1,2)}}}_{\text{\tiny{RPY}}}&
    ...&
    \mathbf{M}^{\text{\tiny{(1,N)}}}_{\text{\tiny{RPY}}}& \\ \mathbf{M}^{\text{\tiny{(2,1)}}}_{\text{\tiny{RPY}}}&
    \mathbf{M}^{\text{\tiny{(2,2)}}}_{\text{\tiny{RPY}}}&
    &
    \mathbf{M}^{\text{\tiny{(2,N)}}}_{\text{\tiny{RPY}}}& \\
    \vdots &
    &
    \ddots &
    \vdots & \\
    \mathbf{M}^{\text{\tiny{(N,1)}}}_{\text{\tiny{RPY}}}&
    \mathbf{M}^{\text{\tiny{(N,2)}}}_{\text{\tiny{RPY}}}&
    ...&
    \mathbf{M}^{\text{\tiny{(N,N)}}}_{\text{\tiny{RPY}}}&
    \end{matrix} \right) = \mathbf{R}^{-1}_{\text{FU}}. 
\end{align*}

In periodic domains, which is the case of our simulations, a compact form of the tensor can be computed in Fourier space, resulting in a positive-definite matrix for any particle configuration~\cite{swan2016rapid,fiore2017rapid,Fiore2018}. The downside of using a RPY-based resistance matrix, is that its pair-wise nature overestimates hydrodynamic forces, when many particles are relatively close. RPY cannot reproduce screening effects at large volume fractions~\cite{durlofsky1987dynamic}, which are known to impact the dynamics in dense suspensions. Gels are partially dense and partially open, which raises the question whether RPY is sufficiently accurate. The Swan group~\cite{vargaswan2015, varga2016hydrodynamic} showed that RPY is able to produce distributions of the particle contact number, that are are representative of those observed in experiment. As we will discuss in Section~\ref{sec:disc}, the contact number distribution may not always be a telling quantifier.

The SD method takes into account far-field, many-body HIs by computing the resistance matrix from a grand-mobility matrix $\mathcal{M}_{\text{ff}}$, \textit{via} matrix inversion. The latter is constructed similarly to $\mathbf{M}_{\text{RPY}}$, but the number of hydrodynamic moments included in the multi-pole expansion is higher than in the RPY formulation~\cite{durlofsky1987dynamic}. As a result, the grand-mobility matrix is an $11N \times 11N$ matrix, and $\mathbf{R}_{\text{FU}}$ is equal to the upper left $6N \times 6N$ block of $\mathcal{M}^{-1}_{\text{ff}}$. The operation of matrix inversion gives rise to many-body hydrodynamic interactions~\cite{durlofsky1987dynamic} (see Fig.~\ref{fig:panel}c), which would be absent otherwise, since $\mathcal{M}_{\text{ff}}$ is computed pair-wise, as with the RPY tensor.

The HIs discussed thus far are all far-field approximations, since the multipole expansion is truncated at a finite order. As a consequence, short-range hydrodynamic forces are poorly reproduced. In particular, the divergent part of such interactions --- following from lubrication --- is completely neglected. In the framework of SD, such interactions are instead computed from the analytical expressions derived by Jeffrey and Onishi~\cite{jeffrey1984calculation} for two-spheres at close proximity. These expressions are used to compute a two-sphere resistance matrix accurate at small separation, which is then used as a building block to assemble the lubrication gran-resistance matrix $\mathcal{R}_{\text{nf}}$, as illustrated in Fig.~\ref{fig:panel}d. However, part of the two-body resistance interactions have already been included upon the inversion of $\mathcal{M}_{\text{{ff}}}$, and therefore, must be subtracted from the total grand-resistance matrix. The matrix containing these interactions is found by simply inverting a two-body mobility matrix. Thus, the total grand-resistance matrix, which contains both near-field lubrication and far-field many-body interactions, is
\begin{align*}
    \mathcal{R} = \mathcal{M}^{-1}_{\text{{ff}}} + \mathcal{R}_{\text{nf}} - \mathcal{R}^{\text{2B}}_{\text{ff}}, 
\end{align*}
where the last term contains the aforementioned corrections. $\mathbf{R}_{\text{FU}}$ is then equal to the upper-left block of $\mathcal{R}$.

It should be noted that here we applied a $1\%$ positive shift in the colloid diameter $\sigma$ used in the inter-particle potential of Eq.~\eqref{eqn:LJpotential}, compared to the one used to construct the resistance matrix $\mathbf{R}_{\text{FU}}$. We did this to prevent particle-particle overlaps, which would make the numerical scheme unstable. Even though lubrication effects are already dominant at surface-to-surface separations $\approx 10\%$ of the colloids diameter, the imposed shift might have weakened the contribution of lubrication to both gel dynamics and structure. Reducing this shift should make the effect of lubrication more prominent.

\section{\label{sec:character}System Characterization}

We used various characterization techniques to understand the dynamics in our colloidal gels and quantify their structure. Here, we predominantly made use of post processing using the Python packages ``freud''~\cite{freud2020} and ``NetworkX''~\cite{SciPyProceedings_11}. In terms of identifying structures, we consider colloids to be bonded, when their center-to-center distance $\Delta r \le 1.05 \sigma$, as in Ref.~\cite{gelhydroJoost2019}.

In order to highlight the effects of HIs, we study the shape of the clusters using the relative shape anisotropy. This is defined as
\begin{align*}
    \langle \kappa^2 \rangle = \frac{3}{2} \left \langle \frac{\sum_{i} \lambda^{4}_i }{(\sum_{i} \lambda^{2}_i )^2}  \right \rangle- \frac{1}{2},
\end{align*}
where the average is performed over the total number of clusters in the system, and $\lambda_i$ are the eigenvalues of each gyration tensor. Values of $\kappa^2$ close to zero identify spherically symmetric clusters, whereas values approaching unity occur only when all the colloids in the cluster lie into a straight line.

To quantify the degree of phase-separation in the system, we make use of the void volume (VV). Here, the VV is computed by considering the volume of a sphere centered at a point in the system, which is just in contact with the surface of the nearest colloid~\cite{haw2006void, koumakis2015tuning, gelhydroJoost2019}. From such a single-point sphere volume, we can compute the average void volume $\langle VV \rangle$ \textit{via} Monte-Carlo integration. The VV allows us to identify structural dissimilarities on the scale of the gel arms, rather than the level of the particles (characterized by the coordination number).

Another quantity capable of indicating the level of coarsening in the gel is the tortuosity parameter $\xi$. This is defined as
\begin{align*}
    \xi = \frac{2}{N_c(N_c - 1)}\sum^{N_c}_{i=1}\sum^{N_c}_{j>i} 1 - \frac{D^{E}(\bold{r}_i,\bold{r}_j)}{D^{N}(\bold{r}_i,\bold{r}_j)},
\end{align*}
with $N_c$ the number of colloids in the considered cluster, $D^{E}$ and $D^{N}$ the Euclidean and network distances respectively. That is, $\xi \in [0,1]$ indicates how tortuous is the gel network, with low values of $\xi$ representing systems close to completing the phase separation, and values close to one for networks with a non-trivial topology and thus, far from completing the phase-separation process. As we will show in Section~\ref{sec:results}, the large differences between the hydrodynamic models studied manifested in $\xi$ rather than VV, with the latter showing smaller deviations.  

\section{\label{sec:results}Results}

In this section, we provide the main results of our study. We start by showing the analogy to the LB-based work by De~Graaf~\cite{gelhydroJoost2019} to demonstrate the various methods provide similar far-field results. Here, we focus mostly on the cluster formation and percolation. Next, we turn to the aging of the system and the way it phase separates. Finally, we consider the state-space trajectories and demonstrate that lubricated interactions lead to a divergent structural pathway.

\subsection{\label{sub:network}Gel Network Formation}

\begin{figure}[!htb]
 \centering
 \includegraphics[width=85mm]{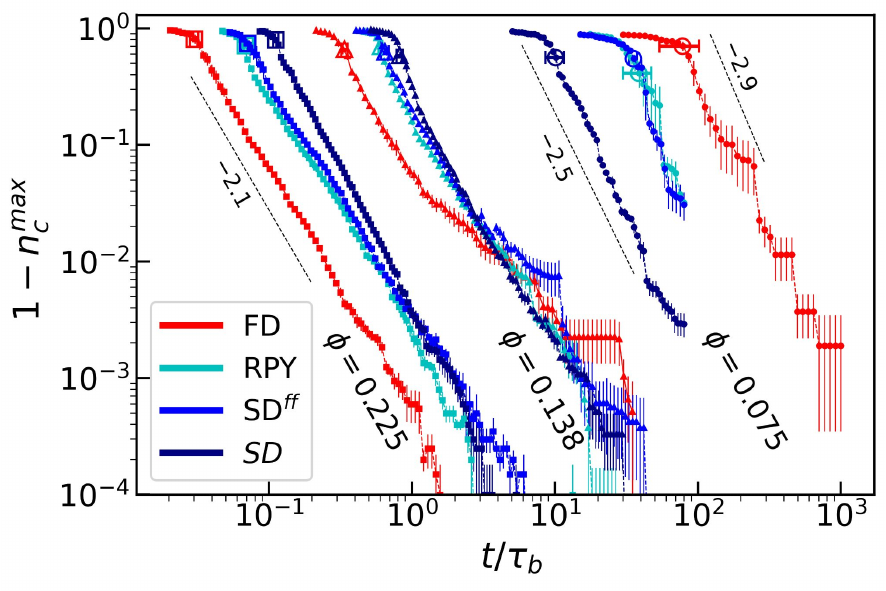}
 \caption{\label{fig:fig1}The fraction of particles in the largest cluster $n_c^{\max}$ (compared to the total number $N$) as a function of time $t$ (in Brownian time units $\tau_{\mathrm{B}}$). We show $1 - n_{c}^{\max}$ to highlight the long-time, power-law scaling for three different colloid volume fractions $\phi$, labelled from small to large using circles, triangles, and squares, respectively. The hydrodynamic approximations are indicated using coloring, as indicated in the legend. Large open symbols indicate the point for which the system percolates. Error bars provide the standard error of the mean. Black dotted lines and numbers indicate the
exponent of asymptotic power laws from Ref.~\cite{gelhydroJoost2019}.}
\end{figure}

We first focus on the short-time dynamics of our systems. Figure~\ref{fig:fig1} provides the fraction of particles $n_c^{\max} = N_c^{\max} / N$ that belong to the largest cluster. We observe that the overall trends are comparable to those of De Graaf~\textit{et al.}~\cite{gelhydroJoost2019}. For the lowest $\phi$, simulations performed with FD exhibit a slower growth than any other hydrodynamic model, while for larger $\phi$, FD predicts the fastest cluster growth; a crossover regime is observed for $\phi = 0.138$. For completeness, we indicate the power-law scaling that was obtained by De Graaf~\textit{et al.}, which appears to match the trends that we find well. Note that particle incorporation in the largest cluster is substantially faster, for our lubricated SD simulations at low $\phi$. We will come back to this result in the following.

\begin{figure}[!htb]
 \centering
 \includegraphics[width=85mm]{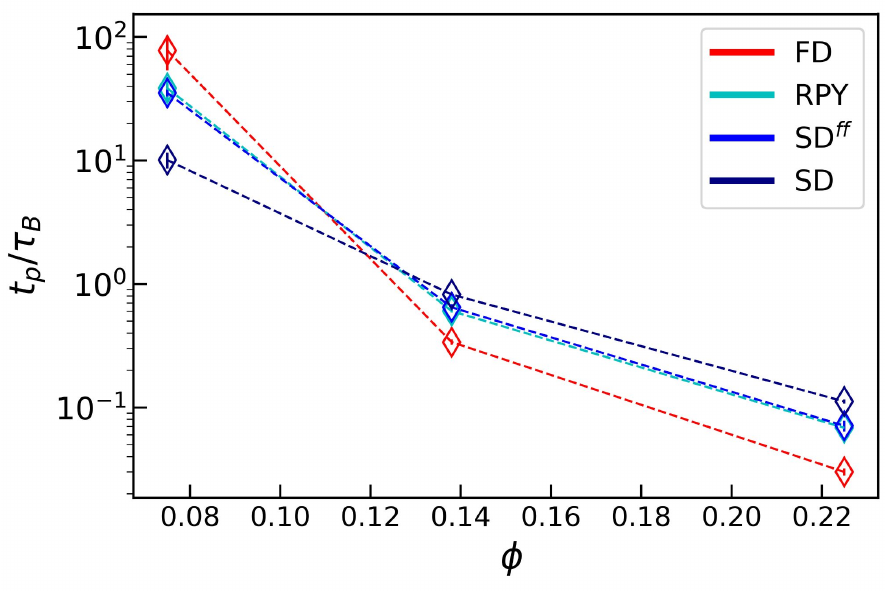}
 \caption{\label{fig:fig2}The percolation time $t_p$ (in terms of $\tau_{\mathrm{B}}$) as a function of $\phi$ for all of our hydrodynamic methods (legend). The dashed lines connect the points belonging to specific method's data set and serve as guides to the eye. The standard error of the mean is indicated using the error bars, but is typically smaller than the symbol size.}
\end{figure}

We also determined the onset of percolation as a function of $\phi$. Figure~\ref{fig:fig2} shows the percolation time $t_p$, defined as the moment at which there exist a cluster that percolates at least one spatial dimension. In line with Ref.~\cite{gelhydroJoost2019}, we find that in all cases, HIs speed up the percolation process at low $\phi$, and slow it down in denser systems. Note that the dramatic effect of lubrication interactions at low $\phi$ is borne out by the percolation transition as well. Contrasting SD and SD$^{\textit{ff}}$ allows us to assess the role of squeeze flows. Our results show that these play a role, but that the effect at intermediate to high volume fractions (for colloidal gelation) is limited. Interestingly, for all regimes, there is limited distinction between RPY and SD$^{\textit{ff}}$ dynamics. This suggests that many-body hydrodynamic effects are of limited importance to gelation.

\begin{figure}[!htb]
 \centering
 \includegraphics[width=85mm]{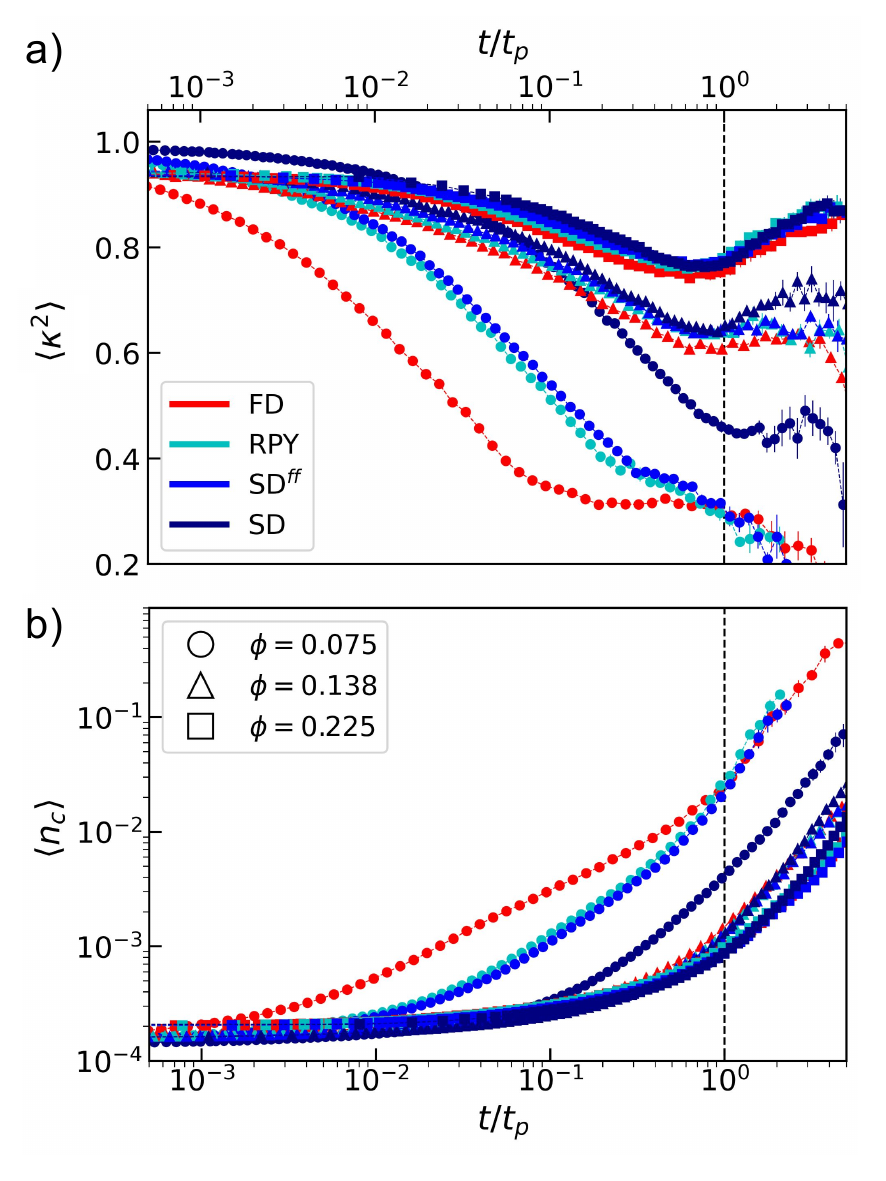}
 \caption{\label{fig:fig3}Features of the short-time clustering in the quenched suspension. (a) The mean relative shape anisotropy $\langle \kappa^2 \rangle$ as a function of time $t$ (in terms of the percolation time $t_p$; emphasized by use of the vertical dashed line). (b) The relative mean cluster size $\langle n_c \rangle = \langle N_c \rangle /N$ also as functions of $t/t_p$. The use of colors, symbols, and error bars is that of Fig.~\ref{fig:fig1}.}
\end{figure}

However, following the largest cluster may not reveal all features of the evolution of the system. We discriminate the effect of various hydrodynamic contributions in the formation of a space-spanning network, by analyzing the shape and size of the clusters that are present in the sample at a given time. Figure~\ref{fig:fig3}a reports the mean relative shape anisotropy, $\langle \kappa^{2} \rangle$. It is clear that this generally decreases as clusters form. This is understood as follows: dimers have $\kappa^{2} = 1$ and, as the clusters grow, the branching structures become more isotropic, leading to a decrease of the overall value of $\langle \kappa^{2} \rangle$.

At short times, newly formed clusters have heterogeneous shapes in any hydrodynamic model considered, and, as they diffuse, these clusters will also reconfigure into more compact shapes in order to minimize their energy. The characteristic time related to the latter process is in competition with percolation, as compact clusters form space-spanning network in a less efficient way compared to more elongated ones. Systems with $\phi \gtrsim 0.138$ show a clear minimum in $\langle \kappa^{2} \rangle$ at the percolation transition. At larger times, the growth of the mean shape anisotropy could be related to the network compaction, as pointed out by Tsurusawa \textit{et al.}~\cite{tsurusawa2019direct} in their experimental study. However, the values of $\kappa^{2}$ are certainly influenced by the periodic boundaries, once a cluster forms that spans the entire simulation box. The long-time increase in $\langle \kappa^2 \rangle$ could be an artifact of this. We deem it irrelevant to our study, as we focus on short-time dynamics ($t \le t_p$) in analyzing the relative shape anisotropy.

Figure~\ref{fig:fig3}b shows the mean number of clustered particles $\langle n_c \rangle = \langle N_c \rangle / N$ as a function of time. It is clear that for intermediate to high $\phi$, the early time cluster dynamics is difficult to distinguish between the various approaches. Systematically, the values for the FD simulations appear lower, but recall that a fair comparison requires an analysis in structure space~\cite{gelhydroJoost2019}. Interestingly, the situation is distinct for low $\phi$, as all curves are initially different, though the FD, RPY, and SD$^{\textit{ff}}$ results start to track each other beyond $t = t_p$. The behavior of our SD simulations always remains separate from these curves, suggesting that lubrication has the ability to modify the structure as well as the dynamics.

In general, the effect of HIs is to both decrease the bonding rate of colloids and increase the time needed to rearrange clusters into lower energy configurations. Thus, on average, HIs impede cluster growth (see Fig.~\ref{fig:fig3}b). For large volume fractions, the time it takes to form a large percolated cluster, which depends on both $\phi$ and colloids bonding rate, is small compared to the time required by clusters to relax their shape. This becomes clear from Fig.~\ref{fig:fig3}a, where large $\phi$ curves overlap once scaled by their respective percolation times. Thus, in that regime of $\phi$, the only noticeable effect of HIs on the gel formation is to decrease the bonding rate of colloids, thereby slowing down short-time gel dynamics.

In contrast, for the the lowest $\phi$, the characteristic time associated with cluster reshaping is comparable with $t_p$, as shown in Fig.~\ref{fig:fig3}a). In this figure, FD systems reach a plateau in the mean relative shape anisotropy $\kappa^2$ before percolation occurs, while the other hydrodynamic models maintain on average more elongated shapes. This promotes the emergence of large structures composed by strand-like clusters of colloids, which speeds up space exploration, the capturing of smaller clusters, and ultimately percolation~\cite{furukawa2010key, royall2015probing, whitmer2011influence}. On the other hand, clusters of particles can rapidly rearrange into more compact configuration in FD simulations. Thus, disfavouring the formation of a percolated network-like structure.

\subsection{\label{sub:aging}Aging and Phase-Separation}

\begin{figure}[!htb]
 \centering
 \includegraphics[width=85mm]{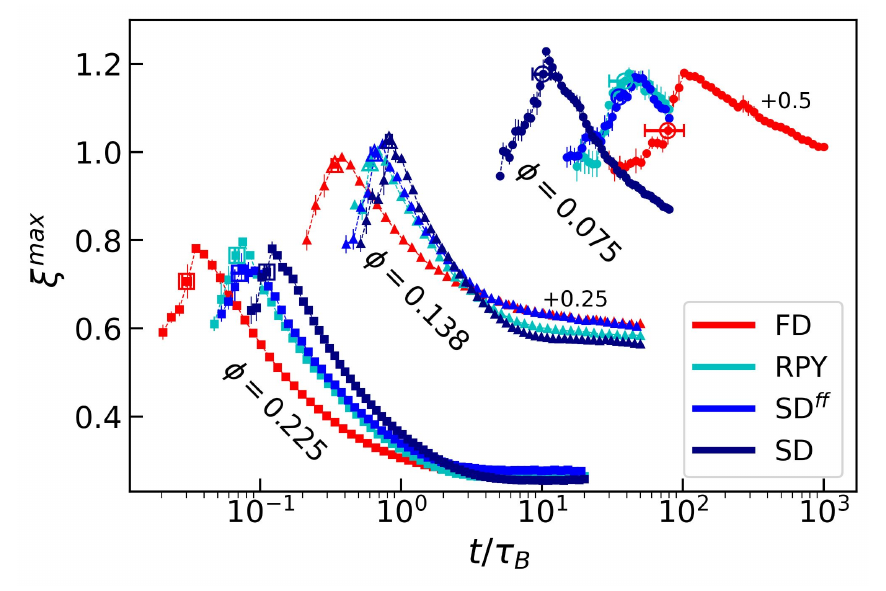}
 \caption{\label{fig:fig4}The tortuosity $\xi$ as a function of time $t$ (in $\tau_{\mathrm{B}}$). The curves for the various values of $\phi$ are shifted (as labelled), in order to better visualize the data. The use of colors, symbols, and error bars is the same as in Fig.~\ref{fig:fig1}.}
\end{figure}

Next, we shift our focus to the long-time dynamics in the gelled system. Figure~\ref{fig:fig4} shows the time evolution of the tortuosity $\xi$ for our three $\phi$ of interest. Note that the curves peak approximately at the respective percolation points. Any mismatch is likely explained by our limited system size and choice of defining percolation as spanning in one direction. After the systems have percolated, the resulting network-like structures will keep spinodally decomposing into colloid-rich and colloid-poor phases, though in an arrested manner. Systems with HIs exhibit a markedly faster decay in $\xi$ for all values of $\phi$ examined, especially when the SD approximation is used. All of the intermediate- to high-$\phi$ systems show a plateauing of $\xi$, suggesting that these display strongly arrested dynamics. Low-$\phi$ systems appear to remain more dynamical over the range that we could simulate. We will argue that this effect is given by the long-range hydrodynamic interactions that propagates among the gel arms.

\begin{figure}[!htb]
 \centering
 \includegraphics[width=85mm]{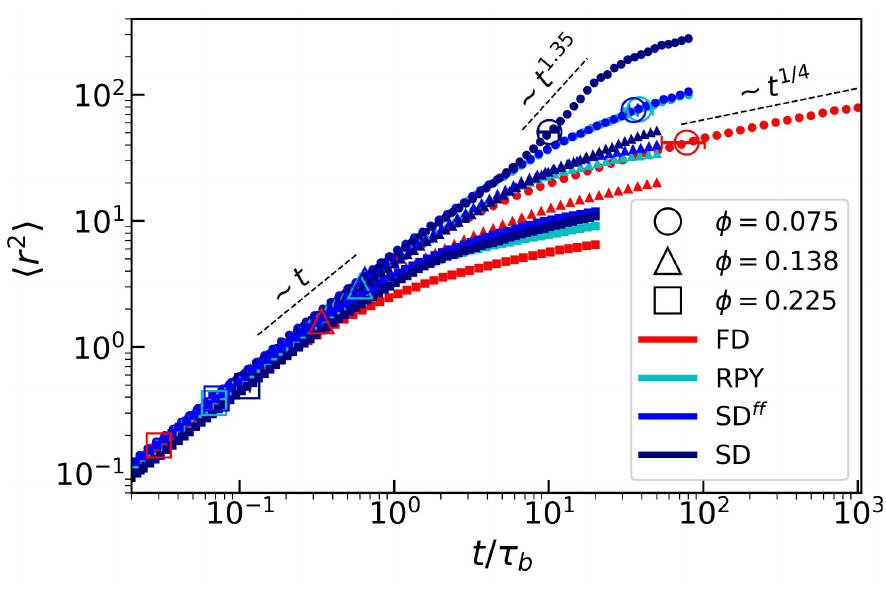}
 \caption{\label{fig:fig5} The mean squared displacement (MSD) $\langle r^{2} \rangle$ averaged over all colloids, expressed as a function of time $t$ (in $\tau_{\mathrm{B}}$). The dashed lines indicate various power-law regimes, while the large symbols indicate the percolation point. The use of colors, symbols, and error bars is the same as in Fig.~\ref{fig:fig1}. However, here, the dashed lines represent fitted power-law trends for the MSD.}
\end{figure}

We also studied the mean squared displacement $\langle r^2 \rangle$ (MSD), averaged over all colloids in the simulation volume. Figure~\ref{fig:fig5} shows that, at short times, all MSDs display a diffusive (linear) regime, as expected for systems that have not yet significantly clustered. In the long-time limit, all systems become subdiffusive, which is representative of the caging colloids experience in the network structure~\cite{wu2020particle}. For our FD system, we were able to simulate sufficiently long to characterize the trend, which appears to follow a power law: $\langle r^2 \rangle \propto t^{1/4}$.

Surprisingly, for the lowest $\phi$, SD-approximated HIs give rise to a transient superdiffusive regime ($\langle r^2 \rangle \propto t^{1.35}$), right after the system percolates. This can be related to the smaller mean cluster size $\langle n_c \rangle$ that SD has before and at percolation, see Fig.~\ref{fig:fig3}b. 

At percolation, the size of the largest cluster is roughly the same in every hydrodynamic model, meaning that in SD simulations the percolated network is surrounded by smaller ``free'' clusters on average. Thermal fluctuations cause displacements of the network arms, which in turn propagate long-range hydrodynamic forces onto the remaining clusters (except in FD systems). Smaller clusters will experience larger displacements due to these interactions, increasing on average the MSD in the SD systems. This is what we alluded to at the end of the opening paragraph to this section. Lastly, all long-time MSDs with HIs are larger than those of the FD systems. The data is suggestive of a collective speeding up of the aging dynamics, which we again relate to HIs generated by the arms of the colloidal gel.  
 
\subsection{\label{sub:state-space}State-Space Trajectories}

\begin{figure}[!htb]
 \centering
 \includegraphics[width=85mm]{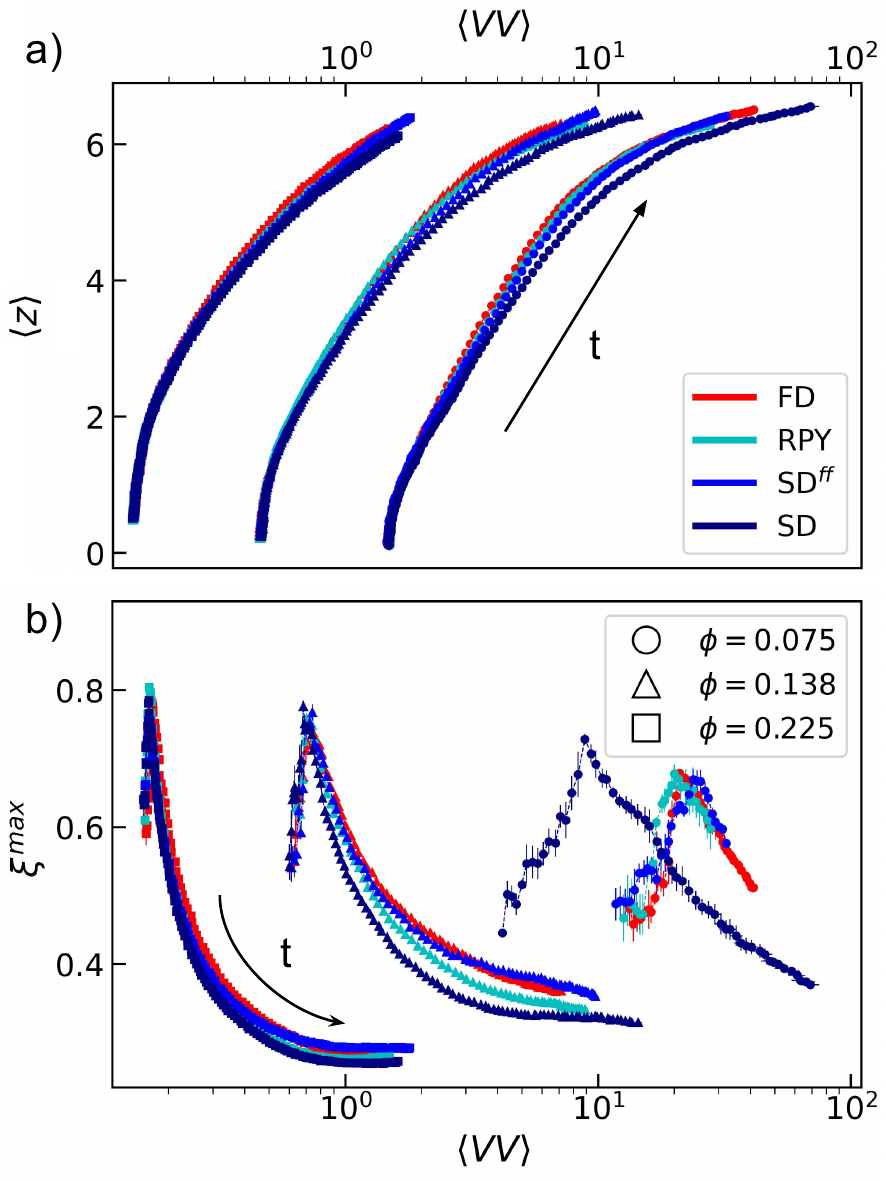}
 \caption{\label{fig:fig6}State-space plots for the structural evolution of colloidal gels. The spaces are spanned by the average void volume $\langle VV \rangle$ and (a) average number of nearest neighbors $\langle z \rangle$ and (b) the tortuosity of the largest cluster $\xi^{\max}$, respectively. The paths are followed in the direction indicated by the black arrow. The use of colors, symbols, and error bars is the same as in Fig.~\ref{fig:fig1}.}
\end{figure}

In the previous section, we have shown that HIs strongly affect the dynamics of the system, which was expected, barring the unexpectedly large impact of HLFs at low $\phi$. Here, we investigate whether this strong change in dynamics is indicative of an actual change in gel structure. Figure~\ref{fig:fig6}a shows the systems' trajectory in a ``state space'' formed by the average void volume $\langle VV \rangle$ and the average number of nearest neighbors $\langle z \rangle$. In this representation, time is parametric and its direction is indicated by the arrow. The results are similar to those reported by De Graaf~\textit{et al.}~\cite{gelhydroJoost2019}. All hydrodynamic models roughly follow the same trajectory in state space, except for systems that include hydrodynamic lubrication. These show (small) deviations that become larger as the $\phi$ is decreased.

We highlight the differences displayed by lubricated systems in Fig.~\ref{fig:fig6}b, which shows the gel trajectories in a state space spanned by $\langle VV \rangle$ and the tortuosity of the largest cluster $\xi^{\max}$. The qualitative picture remain roughly unchanged for large and intermediate $\phi$, with deviations from the simple FD predictions becoming more significant as the hydrodynamic model considered becomes more accurate. For the lowest $\phi$ considered here, SD systems pursue a clearly distinct path in state space from those obtained using FD, RPY, and SD$^{\textit{ff}}$. We conclude that tortuosity is a better characterizer of differences in colloidal gels than void volumes, and that the HLFs can strongly impact structure at low $\phi$.  

\section{\label{sec:disc}Discussion}

We showed that far-field HIs change the dynamics of colloidal gel formation, as already found in other numerical studies of gelation~\cite{furukawa2010key, whitmer2011influence, royall2015probing, vargaswan2015, varga2016hydrodynamic, gelhydroJoost2019}. However, unless HLFs are included, this has little impact on the structural paths that these systems explore as they form, in line with De~Graaf~\textit{et al.}~\cite{gelhydroJoost2019}. Nonetheless, HLFs can strongly impact the dynamics and structure of a gel, in contradiction to the results of Bybee and Higdon~\cite{bybee2009hydrodynamic}. The important difference here is that we include far-field effects. The strong effect of HLFs on structure are thus likely a consequence of an interplay with far-field HIs. Speculating, it could be that the fluid particle dynamics simulations by Furukawa and Tanaka~\cite{furukawa2010key} picked up on this effect, as these authors used relatively low volume fractions, weak gels, and numerically refined particles.

Let us now explain the counter intuitive result obtained by combining HLFs and far-field HIs during aging. The effect of far-field-only HIs on the aging process is clear: a less static network will coarsen faster over time. Adding HFLs to this, accelerates aging further, for two reasons: (i)~Divergent dissipative forces hinder both the creation and rupture of colloidal bonds. That is, it is harder to push the colloids in and out of contact, meaning that once arms have formed, they are less likely to break. (ii)~Thermal fluctuations are more collective in nature, when including HLFs. To understand this, note that Brownian velocities can be decomposed into a near- and a far-field contribution~\cite{fiore2019fast}: $\mathcal{F^B} = \mathcal{F^B}_{\text{nf}} + \mathcal{F^B}_{\text{ff}}$. The former are fundamentally different when lubrication is included. In this case, the effect of lubrication is to filter out relative thermal displacements and rotations of colloids that are in close proximity. Thus, the spectrum of fluctuations is mainly populated by collective modes, which displace coherently the colloids that belong to the same gel arm and facilitates the coarsening process. As a result, the flux of particles migrating from the colloid-rich to the colloid-poor regions is negatively affected by HLFs, thus, favouring phase-separation. It would be worthwhile to revisit the work on the hopping dynamics of colloids in a gel arm~\textit{et al.}~\cite{van2018strand, verweij2019plasticity} in view of this.

In terms of computational cost, SD simulations are clearly the most expensive. This cost stems mainly from the matrix inversion needed to reproduce many-body far-field HIs, rather than the inclusion of near-field lubrication. Fortunately, HLFs slow down bonding dynamics, allowing the use of larger $\Delta t$, which makes SD effectively more efficient than SD$^{\textit{ff}}$. Nonetheless, full SD remains a costly numerical scheme in systems with strong short-ranged interactions such as colloidal gels, practically limiting the system sizes that can be studied to a few thousand particles. Considering that RPY and SD$^{\textit{ff}}$ have very similar behaviors, it could be worthwhile to investigate how accurate a lubricated RPY algorithm is in reproducing full SD results. If this approach is successful, much larger system sizes can be considered.

\section{\label{sec:conclusion}Conclusions and Outlook}

Summarizing, we have studied the effect of hydrodynamic interactions in the formation and aging of colloidal gels. We considered single-body (Brownian Dynamics), pair-wise (Rotne-Prager-Yamakawa), many-body (Stokesian Dynamics), and lubrication-corrected many-body interactions. As expected, we found that the dynamics of gel formation is sensitive to the exact nature of the approximation that is used. However, the steady-state structure is relatively unaffected when considering far-field hydrodynamic interactions and appropriate rescaling of time, in line with a previous analysis based on lattice Boltzmann simulations.

Intriguingly, introducing hydrodynamic lubrication corrections results in significant departures from the rescaled trend, especially for colloid volume fractions $\phi \lesssim 0.138$. Not only is the structure altered, with on average smaller voids and clusters at the percolation point, lubrication also accelerates the aging of the gel. This result seems counterintuitive, as the lubricated regime is typically associated with the strongest hydrodynamic dissipation, and is not in line with previous understanding in the literature. Both the dynamic and structure effect can, however, be explained by the fact that lubrication interactions hinder the bonding and rupturing of clusters, as well as suppress non-collective Brownian modes in the gel arms. These aspects combine to facilitate phase-separation. The key point this that both far-field hydrodynamics and lubrication forces must be present to realize is enhanced separation, which is where our study improves upon the existing literature. Studying colloidal gelation thus requires a relatively costly, but accurate representation of hydrodynamic interactions.

Moving away from the idealized Stokes-flow (approximate) solutions that we considered here, our findings strongly suggests that near-contact dynamics should be given due consideration in the modeling of experimental colloidal gels. This work provides a solid foundation for future studies that make the comparison between experiment and simulation. In particular, it will be relevant to see how the lubricated dynamics impacts gels subjected to externally imposed stimuli, such as shear.\\

\section*{Acknowledgements}

The authors acknowledge NWO for funding through OCENW.KLEIN.354. We are grateful to the late Prof.~James Swan for initial discussions on the possible differences between LB and SD simulations of colloidal gels. We thank Dr. Andrew Fiore and Dr. Madhu Majji for their help with getting PSE up and running with HOOMD-blue, as well as Dr. Gwynn J. Elfring and Dr. Zhouyang Ge for sharing an accelerated variant of PSE and discussing minor bugs in the resistance tensor. An open data package containing the means to reproduce the results of the simulations is available at: [DOI]

\bibliographystyle{aip}
\bibliography{reference}

\end{document}